\documentclass[english]{article}
\usepackage[T1]{fontenc}
\usepackage[latin9]{inputenc}
\usepackage{geometry}
\usepackage{amsmath}
\usepackage{amssymb}
\usepackage{graphicx}
\usepackage{babel}
\begin{document}

\title{Embarrassingly Parallel Time Series Analysis for\\
Large Scale Weak Memory Systems}

\author{Francois Belletti, Evan Sparks, Michael Franklin, Alexandre M. Bayen}
\maketitle
\begin{abstract}
Second order stationary models in time series analysis are based on
the analysis of essential statistics whose computations follow a common
pattern. In particular, with a map-reduce nomenclature, most of these
operations can be modeled as mapping a kernel that only depends on
short windows of consecutive data and reducing the results produced
by each computation. This computational pattern stems from the ergodicity
of the model under consideration and is often referred to as weak
or short memory when it comes to data indexed with respect to time.
In the following we will show how studying weak memory systems can
be done in a scalable manner thanks to a framework relying on specifically
designed overlapping distributed data structures that enable fragmentation
and replication of the data across many machines as well as parallelism
in computations. This scheme has been implemented for Apache Spark but is
certainly not system specific. Indeed we prove it is also adapted
to leveraging high bandwidth fragmented memory blocks on GPUs.

\end{abstract}

\section*{Introduction}

Classic time series analysis and in particular that of second order
stationary processes has been widely studied for several decades and
reached a peak of popularity in the years 1980-1990 when computational
power became more common and allowed for practical implementation
of varied data sets. Monographs such as \cite{brillinger1981time},
\cite{Brockwell:1986:TST:17326} and \cite{hamilton1994time} offered
the opportunity for many researchers to better understand the intricacies
of time series analysis. The topics of univariate and monovariate
time series analysis were also well covered in \cite{harvey1993time,Ltkepohl:2007:NIM:1554948}
and many other publications. Software such as R or Matlab has given
the opportunity to many practitioners to examine data, enabling even
more applications of time series analysis and contributing to the
interest in that field. The amount of research dedicated to time series
analysis has been so important in the past decades that we do not
claim here to review it holistically. However, it is necessary to
contextualize the topic of the present article both in its theoretical
aspects and its practical implementation.\\
The development of distributed programming paradigms \cite{dean2008mapreduce},
file systems \cite{shvachko2010hadoop}, databases \cite{thusoo2010hive}
and in-memory computing engines \cite{zaharia2012resilient} 
for the modern commodity datacenter environment has led
us to consider new applications and implementations for time series
analysis. As opposed to small dataset applications such as those developed
in quantitative finance \cite{tsay2005analysis}, climate studies
\cite{mudelsee2010climate}, network traffic analysis \cite{basu1996time},
hydrology \cite{hipel1994time}, we are now interested in analyzing
observations at scale in a data intensive environment such as \cite{hunter2011scaling}
with a standardized distributed library similar to MLlib \cite{franklin2013mllib}.
This creates new challenges. In particular, we are dealing with data
sitting on a distributed cluster of machines whose memory layout has
an obvious bandwidth bottleneck when it comes to shuffling data across
an Ethernet network. This also creates new opportunities. Time series
estimation can now leverage recent theoretical developments
\cite{durbin2012time} as well as improvements in convex optimization
techniques \cite{boyd2004convex,bertsekas1999nonlinear} that can
help get likelihood maximization based estimators faster.\\
In this new programming framework, we will start adapting the implementation
of the simplest class of models: linear time series models. These are the stochastic
counterparts of deterministic linear systems \cite{callier2012linear}.
Estimating auto-regressive and moving average models (weak memory
models) will be the main focus of this document. More advanced topics
such as the identification of non linear dynamics \cite{kantz2004nonlinear,fan2003nonlinear}
as well heteroscedastic systems \cite{Straumann05estimationin} will
be the subject of subsequent work.\\
This document focuses on the presentation of a new data structure
dedicated to the representation of estimation-ready data sets for
time series analysis. As opposed to other frameworks such as SparkTS,
spark-timeseries or Thunder, the data is partitioned here with respect
to time. An overlapping block distributed data structure has been
devised which, given an appropriate padding expressed in units of
time, enables the computation of M and Z estimators for second order
stationary models in an embarrassingly parallel manner. This enables
in particular the calibration of multivariate time series models without
shuffling observations on a distributed in-memory computing engine.

\part{Theoretical background on time series analysis, estimators, motivation}

Our intent is to provide a generic programming framework for scalable
time series analysis in a distributed system. In the following, we
review popular time series models that should be supported by the
framework and establish the corresponding computational operations
the related estimators require. Focus is set on second order stationary
models and the the well known ARMA family.

\section{Second order stationary time series}

We start with the most common time series models. These are models
in which observation are regularly spaced and for which there are
no missing values. Practitioners of time series analysis are familiar
with missing data, outliers and irregularly spaced timestamps. In
such cases, an interpolation technique (linear interpolation or last-observation-carried-forward
for instance) is often used in order to align observations on a regular
time index grid. We focus on time series $\left(X_{t}\right)_{t\in\mathbb{Z}}$
where each observation belongs in $\mathbb{R}^{d}$.

\subsection{Second order stationarity}

Second order stationary processes are common in time series analysis.
Models in this family are simple, practical to estimate and yet have
strong predictive power for a vast range of data sets related to economics,
finance, industrial systems, data center monitoring and the climate.

\paragraph{Definition: Second order stationary time series}

A time series $\left(X\right)=\left(X_{t}\right)_{t\in\mathbb{Z}}$
is second order stationary if there exists an auto-covariance function
$\gamma:\mathbb{Z}\rightarrow\mathbb{R}^{d\times d}$ so that for
any value of $t$, 
\[
\mbox{Cov}\left(X_{t},X_{t+h}\right)=\gamma^{X}\left(h\right).
\]
White noise is an example of second order stationary time series where
the auto-covariance function is 0 everywhere except for $h=0$.

\paragraph*{Definition: lag operator $L$}

Let $L:\left(\mathbb{R}^{d}\right)^{\mathbb{Z}}\rightarrow\left(\mathbb{R}^{d}\right)^{\mathbb{Z}}$
the operator such that $L\left(\left(X_{t}\right)_{t\in\mathbb{Z}}\right)=\left(X_{t-1}\right)_{t\in\mathbb{Z}}$.

\subsection{Constant volatility linear time series models}

A first family of models is concerned with modeling observations as
the output of a linear system with constant variance perturbations.

\paragraph{Definition: multidimensional white noise}

A process $\left(\varepsilon_{t}\right)_{t\in\mathbb{Z}}$ in $\mathbb{R}^{d}$
is a multidimensional white noise if 
\[
\forall t\in\mathbb{Z},\: E\left(\varepsilon_{t}\right)=0
\]
\[
\forall t\in\mathbb{Z},\: E\left(\varepsilon_{t}\varepsilon_{t}^{T}\right)=\Sigma_{\varepsilon}
\]
and 
\[
\forall t,s\in\mathbb{Z}\::\: t\neq s,\: E\left(\varepsilon_{t}\varepsilon_{s}^{T}\right)=0.
\]
In the following we assume white noise processes are always non-degenerate,
i.e., we always have $E\left(\varepsilon_{t}\varepsilon_{t}^{T}\right)=\Sigma_{\varepsilon}$
definite positive.

\paragraph{Definition: Auto-regressive models (AR)}

A (centered) multivariate order $p$ auto-regressive time series is
defined by $p$ matrices $\left(A_{k}\right)_{k\in\left\{ 1\ldots p\right\} }\in\mathbb{R}^{d\times d}$
and a white noise process $\left(\varepsilon_{t}\right)_{t\in\mathbb{Z}}$
in $\mathbb{R}^{d}$ with variance $\Sigma_{\varepsilon}\in\left(\mathbb{R}^{d\times d}\right)$
such that, for any value of $t$ in $\mathbb{Z}$, 
\[
X_{t}=A_{1}X_{t-1}+\ldots+A_{p}X_{t-p}+\varepsilon_{t}.
\]
This equation can be rewritten in reduced form with the lag operator.
Let $A\left(z\right)=I-A_{1}z-\ldots-A_{p}z^{p}$ be the companion
polynomial of the equation. A short hand for the AR equation above
is 
\[
A\left(L\right)X=\varepsilon.
\]
Equivalently, a Linear Time Invariant (LTI) system formulation of
these equations is: 
\[
\left(\begin{array}{c}
X_{t}\\
X_{t-1}\\
\vdots\\
X_{t-p+1}
\end{array}\right)=\left(\begin{array}{ccccc}
A_{1} & A_{2} & \cdots & \cdots & A_{p}\\
I_{d} & 0\\
 & I_{d} & \ddots\\
 &  & \ddots & \ddots\\
 &  &  & I_{d} & 0
\end{array}\right)\left(\begin{array}{c}
X_{t-1}\\
X_{t-2}\\
\vdots\\
X_{t-p}
\end{array}\right)+\left(\begin{array}{c}
\varepsilon_{t}\\
0\\
\vdots\\
0
\end{array}\right).
\]
In the following we will mostly use the first AR formulation as the
LTI formulation has a degenerate noise structure (the noise covariance
matrix is not full rank).

\paragraph{Definition: Moving average models (MA)}

A (centered) multivariate order $q$ moving average time series is
defined by $q$ matrices $\left(B_{k}\right)_{k\in\left\{ 1\ldots q\right\} }\in\mathbb{R}^{d\times d}$
and a white noise process $\left(\varepsilon_{t}\right)_{t\in\mathbb{Z}}$
in $\mathbb{R}^{d}$ with variance $\Sigma_{\varepsilon}\in\left(\mathbb{R}^{d\times d}\right)$
such that, for any value of $t$ in $\mathbb{Z}$, 
\[
X_{t}=\varepsilon_{t}+B_{1}\varepsilon_{t-1}+\ldots+B_{q}\varepsilon_{t-q}.
\]
Letting $B\left(z\right)=I+B_{1}z-\ldots+B_{q}z^{q}$ the companion
polynomial of the equation. A short hand for the MA equation above
is 
\[
B\left(L\right)X=\varepsilon.
\]

\paragraph{Definition: Auto-regressive moving average models (ARMA)}

A (centered) multivariate order $p$, $q$ auto-regressive moving average
time series is defined by $p$ matrices $\left(A_{k}\right)_{k\in\left\{ 1\ldots p\right\} }\in\mathbb{R}^{d\times d}$
and $q$ matrices $\left(B_{k}\right)_{k\in\left\{ 1\ldots q\right\} }\in\mathbb{R}^{d\times d}$
and a white noise process $\left(\varepsilon_{t}\right)_{t\in\mathbb{Z}}$
in $\mathbb{R}^{d}$ with variance $\Sigma_{\varepsilon}\in diag\left(\mathbb{R}^{d\times d}\right)$
such that, for any value of $t$ in $\mathbb{Z}$, 
\[
X_{t}=A_{1}X_{t-1}+\ldots+A_{p}X_{t-p}+\varepsilon_{t}+B_{1}\varepsilon_{t-1}+\ldots+B_{q}\varepsilon_{t-q}.
\]
Such a model corresponds to a perturbed LTI system whose perturbations
are auto-correlated.\\
\\
Letting $A\left(z\right)=I-A_{1}z-\ldots-A_{p}z^{p}$ and $B\left(z\right)=I+B_{1}z-\ldots+B_{q}z^{q}$
the companion polynomials of the equation. A short hand for the ARMA
equation above is 
\[
A\left(B\right)X=B\left(L\right)\varepsilon.
\]

\subsection{Identification, causality and invertibility}

In the following, we will consider the conditions that guarantee stability
of the model and the fact that the equations above have one and only
solution.

\subsubsection{Causal models:}

If $\forall z:\:\left|z\right|<1,\: det\left(A\left(z\right)\right)\neq0$
i.e. if the matrix 
\[
\left(\begin{array}{ccccc}
A_{1} & A_{2} & \cdots & \cdots & A_{p}\\
I_{d} & 0\\
 & I_{d} & \ddots\\
 &  & \ddots & \ddots\\
 &  &  & I_{d} & 0
\end{array}\right)
\]
has its spectrum strictly bounded in absolute value by $1$, the ARMA
equation has one and only stationary solution. This solution is said
to be causal. Such an ARMA equation can be interpreted as equivalent
to a stable LTI which is perturbed by auto-correlated noise.\\
In the following we will only consider causal models.

\subsubsection{Invertible models:}

If $\forall z:\:\left|z\right|<1,\: det\left(B\left(z\right)\right)\neq0$
i.e. if the matrix 
\[
\left(\begin{array}{ccccc}
-B_{1} & -B_{2} & \cdots & \cdots & -B_{q}\\
I_{d} & 0\\
 & I_{d} & \ddots\\
 &  & \ddots & \ddots\\
 &  &  & I_{d} & 0
\end{array}\right)
\]
has its spectrum strictly bounded in absolute value by $1$, for a
stationary solution to the ARMA equations $\left(X\right)$ there
exists a unique white noise process $\left(\varepsilon\right)$ such
that $A\left(L\right)X=B\left(L\right)\varepsilon$.

\subsection{Integrated processes and differentiation}

\paragraph{Definition: differentiated time series}

For any time series $\left(X_{t}\right)_{t\in\mathbb{Z}}$ one may
define its differentiated counterpart as $\triangle\left(\left(X_{t}\right)_{t\in\mathbb{Z}}\right)=\left(X_{t+1}-X_{t}\right)_{t\in\mathbb{Z}}$.\\
In the case of actual and finite length data, we opt for the convention:
$\triangle\left(\left(X_{t}\right)_{t\in\left\{ 1\ldots N\right\} }\right)=\left(X_{t+1}-X_{t}\right)_{t\in\left\{ 1\ldots N-1\right\} }$.

\paragraph{Definition: integrated processes}

Formally, a time series $\left(X_{t}\right)_{t\in\mathbb{Z}}$ is
said to be integrated of order $I$ if $\triangle^{d}\left(X\right)$
is not second order stationary whenever $d<I$ and $\triangle^{I}\left(X\right)$
is second order stationary. \\
\\
Brownian motions are famous examples of order $1$ integrated processes
(there difference process is a white noise).

\section{Estimators of sufficient statistics of second order stationary time
series}

In the following we review known estimators for multivariate time
series in order to highlight the similarity of their computational
structure. We are concerned with a theoretical process $\left(X\right)$
and have $\left(X_{t}\right)_{t\in\left\{ 1\ldots N\right\} }$ consecutive
observations.

\subsection{Estimators of interest}

The following section goes through all the sufficient statistics to
estimate a second order stationary process. In particular, it is noteworthy
that the causal solution to an ARMA equation has a covariance structure
that is entirely determined by the parameters of the equation \cite{Brockwell:1986:TST:17326}
(that is $A_{1},\ldots,A_{p}$, $B_{1},\ldots,B_{q}$ and $\Sigma$).

\subsubsection{Mean}

A consistent unbiased estimator for the mean of a second order time
series $\left(X\right)$ is 
\[
\widehat{\mu^{X}}\left(\left(X_{t}\right)_{t\in\left\{ 1\ldots N\right\} }\right)=\frac{1}{N}\sum_{k=1}^{N}X_{k}.
\]
This estimator is asymptotically normal with variance decaying with
$\frac{1}{N}$ rate. \\
\\
In the following we assume that the mean has been estimated and accounted for
and $\left(X\right)$ is therefore centered.

\subsubsection{Auto-covariance}

A consistent unbiased estimator for the auto-covariance matrix at
lag $h\in\mathbb{Z}$ for a centered second order stationary time
series is 
\[
\widehat{\gamma^{X}\left(h\right)}\left(\left(X_{t}\right)_{t\in\left\{ 1\ldots N\right\} }\right)=\frac{1}{N-h-1}\sum_{k=1}^{N-h}X_{k}X_{k+h}^{T}.
\]
This estimator is asymptotically normal with variance decaying with
$\frac{1}{N}$ rate.

\subsubsection{Auto-correlogram}

\paragraph{Definition: auto-correlogram of a second order stationary time series}

Let $h\in\mathbb{Z}$, order $h$ auto-correlation is 

\[
\rho_{h}^{X}=\mbox{Cor}\left(X_{t},X_{t+h}\right)=\mbox{diag}\left(\gamma^{X}\left(0\right)\right)^{-\frac{1}{2}}\gamma^{X}\left(h\right)\mbox{diag}\left(\gamma^{X}\left(0\right)\right)^{-\frac{1}{2}}.
\]

\paragraph*{Auto-correlation estimator:}

The estimator

\[
\widehat{\rho_{h}^{X}}=\mbox{diag}\left(\widehat{\gamma^{X}}\left(0\right)\right)^{-\frac{1}{2}}\widehat{\gamma^{X}\left(h\right)}\mbox{diag}\left(\widehat{\gamma^{X}\left(0\right)}\right)^{-\frac{1}{2}}
\]
is asymptotically convergent, it is in fact asymptotically normal
with variance decaying with $\frac{1}{N}$ rate.

\paragraph{Definition: partial auto-correlogram of a second order stationary
time series}

Let $H_{X}^{t-p,t}$ the span of the random variables $\left\{ X_{t-p},\:\ldots,\: X_{t}\right\} $
and $P_{H_{X}^{p,t}}$ the corresponding orthogonal projection. Let
$\left(U_{j}^{\left(p\right)}\right)_{j=1}^{p}$ be defined by 
\[
P_{H_{X}^{t-p,t-1}}X_{t}=\sum_{j=1}^{p}U_{j}^{\left(p\right)}X_{t-j}.
\]
$U_{p}^{\left(p\right)}$ is the partial auto-correlation matrix of
order $p$, we will denote it $\kappa^{X}\left(p\right)$.

\paragraph{Property: from auto-correlation to partial auto-correlation}

The projection matrices above solve the following linear system of
equations: 
\[
\left[\begin{array}{ccc}
\gamma^{X}\left(0\right) & \cdots & \gamma^{X}\left(-\left(p-1\right)\right)\\
\vdots & \ddots & \vdots\\
\gamma^{X}\left(p-1\right) & \cdots & \gamma^{X}\left(0\right)
\end{array}\right]\left[\begin{array}{c}
\left(U_{1}^{\left(p\right)}\right)^{T}\\
\vdots\\
\left(U_{p}^{\left(p\right)}\right)^{T}
\end{array}\right]=\left[\begin{array}{c}
\gamma^{X}\left(1\right)\\
\vdots\\
\gamma^{X}\left(p\right)
\end{array}\right].
\]
\textbf{Proof: }

$P_{H_{X}^{t-p,t-1}}$ is an orthogonal projector therefore for any
$j\in\left\{ 1\ldots p\right\} $, 
\[
E\left[\left(X_{t}-P_{H_{X}^{t-p,t-1}}X_{t}\right)X_{t-j}^{T}\right]=0
\]
Therefore, 
\[
E\left[\left(\begin{array}{c}
X_{t-1}\\
\vdots\\
X_{t-p}
\end{array}\right)\left(X_{t}\right)^{T}\right]=E\left[\left(\begin{array}{c}
X_{t-1}\\
\vdots\\
X_{t-p}
\end{array}\right)\left(P_{H_{X}^{t-p,t-1}}X_{t}\right)^{T}\right]=E\left[\left(\begin{array}{c}
X_{t-1}\\
\vdots\\
X_{t-p}
\end{array}\right)\left(\begin{array}{ccc}
X_{t-1}^{T} & \ldots & X_{t-p}^{T}\end{array}\right)\left(\begin{array}{c}
\left(U_{1}^{\left(p\right)}\right)^{T}\\
\vdots\\
\left(U_{p}^{\left(p\right)}\right)^{T}
\end{array}\right)\right].
\]

\paragraph{Partial auto-correlation estimator:}

An estimator $\widehat{\kappa^{X}\left(p\right)}$ can be obtained
by inverting the linear system above where the actual auto-covariance
is replaced by $\widehat{\gamma}$:

\[
\left[\begin{array}{ccc}
\widehat{\gamma^{X}}\left(0\right) & \cdots & \widehat{\gamma^{X}}\left(-\left(p-1\right)\right)\\
\vdots & \ddots & \vdots\\
\widehat{\gamma^{X}}\left(p-1\right) & \cdots & \widehat{\gamma^{X}}\left(0\right)
\end{array}\right]\left[\begin{array}{c}
\vdots\\
\vdots\\
\left(\widehat{\kappa^{X}\left(p\right)}\right)^{T}
\end{array}\right]=\left[\begin{array}{c}
\widehat{\gamma^{X}}\left(1\right)\\
\vdots\\
\widehat{\gamma^{X}}\left(p\right)
\end{array}\right].
\]
It is asymptotically unbiased and asymptotically normal with variance
decaying with a $\frac{1}{N}$ rate.

\section{AR, MA, ARMA fitting by frequentist methods}

In the following we assume $\left(X\right)$ is a centered second
order stationary process.

\subsection{Determining the order of an AR, MA or ARMA model by frequentist methods}

It is possible to choose an appropriate value of $p$ when estimating
an AR model simply by computing the partial auto-correlation of the
process. As soon as the partial auto-correlation at lag $h$ is not
significantly different from $0$, an appropriate choice for $p$
is $h-1$. Indeed, for an AR model of order $p$, partial auto-correlation
$\kappa^{X}\left(h\right)$ cancels out as soon as $h>p$.\\
\\
Similarly, if one considers a MA model of order $q$, the auto-correlation
function $\gamma^{X}\left(h\right)$ is zero whenever $h>q$. Therefore,
value of $h$ after which the auto-correlation function is no longer
significantly different from zero yields an indicator of the value
of $q$ one should choose prior to estimating the model.\\
\\
For ARMA models, the analysis is more involved but only relies on
estimates of auto-covariance as well. In this case, it is more common
in practice to choose cutoff values for $p$ and $q$ based on a Bayesian
information criterion (AIC or BIC).

\subsection{AR estimation based on Yule-Walker equations:}

Assuming 
\[
X_{t}=A_{1}X_{t-1}+\ldots+A_{p}X_{t-p}+\varepsilon_{t}
\]
the Yule-Walker equations give 
\[
\left[\begin{array}{ccc}
\gamma^{X}\left(0\right) & \cdots & \gamma^{X}\left(-\left(p-1\right)\right)\\
\vdots & \ddots & \vdots\\
\gamma^{X}\left(p-1\right) & \cdots & \gamma^{X}\left(0\right)
\end{array}\right]\left[\begin{array}{c}
A_{1}^{T}\\
\vdots\\
A_{p}^{T}
\end{array}\right]=\left[\begin{array}{c}
\gamma^{X}\left(1\right)\\
\vdots\\
\gamma^{X}\left(p\right)
\end{array}\right]
\]
or equivalently 
\[
\left[\begin{array}{ccc}
A_{1} & \cdots & A_{p}\end{array}\right]\left[\begin{array}{ccc}
\left(\gamma^{X}\left(0\right)\right)^{T} & \cdots & \left(\gamma^{X}\left(p-1\right)\right)^{T}\\
\vdots & \ddots & \vdots\\
\left(\gamma^{X}\left(-\left(p-1\right)\right)\right)^{T} & \cdots & \left(\gamma^{X}\left(0\right)\right)^{T}
\end{array}\right]=\left[\begin{array}{ccc}
\left(\gamma^{X}\left(1\right)\right)^{T} & \cdots & \left(\gamma^{X}\left(p\right)\right)^{T}\end{array}\right].
\]
Solving this block Toeplitz linear system with the auto-covariance
estimators given above then yields the least square estimates of $A_{1}^{T},\ldots,A_{p}^{T}$.
Right multiplying the equation above by $X_{t}^{T}$ and computing
the expectation gives an estimate of the diagonal variance matrix
of the noise process:
\[
\Sigma_{\varepsilon}=\gamma_{0}^{X}-A_{1}\gamma_{-1}^{X}-\ldots-A_{p}\gamma_{-p}^{X}
\]
The interesting point here, from a computational standpoint, is the
fact that auto-covariance estimates for $h=0\ldots p-1$ are sufficient
to obtain estimates for the parameters of the model. In the univariate
case this comes down to inverting a Toeplitz matrix and is practically
achieved thanks to the well known Durbin-Levinson algorithm \cite{Brockwell:1986:TST:17326}
with $O\left(p^{2}\right)$ time complexity.\\
\\
The multivariate case is more in-line with the kind of large scale
analytics programming paradigm we discuss here. Indeed, what is key
to the present approach is to be able to fit AR models not specifically
with a large order $p$ but more with a large number of dimensions
$d$. One is therefore interested in solving such a system with large
Toeplitz blocks (dimension $d$) and small order in comparison $\left(p\ll d\right)$.
Akaike offered a recursive method to solve such a system with $O\left(p^{2}\times d\right)$
extra space and $O\left(p^{2}\times d^{3}\right)$ in \cite{akaike}.

\subsubsection{Issues when $d$ becomes very large:}

The fact that, in this algorithm one has to invert matrices of size
$\left(d,\: d\right)$ becomes problematic whenever $d$ becomes large
($10^{5}$ or more). The time complexity of the block Toeplitz inversion 
procedure of Akaike is cubic with respect to $d$ which means practically
that even on modern GPUs capable of 1 TFLOPs, 
it would generally take at least $12$ days to invert a $10^6$ row square matrix.
Therefore we show how to conduct multivariate
analysis when $d$ is high and the system under study features spatial
stationarity thanks to a Bayesian approach.

\subsection{MA estimation based on the innovation algorithm:}

We want to estimate a model of the form 
\[
X_{t}=\varepsilon_{t}+B_{1}\varepsilon_{t-1}+\ldots+B_{q}\varepsilon_{t-q}
\]
Let us assume that their exist $\left(\Theta_{m,n}\right)_{m,n\in\left\{ 1\ldots q\right\} }$
such that for for any $m\in\left\{ 1\ldots q\right\} $. 
\[
P_{H_{X}^{1,m}}X_{m+1}=\begin{cases}
0 & \mbox{if}\: m=0\\
\sum_{j=1}^{m}\Theta_{m,j}\left(X_{m-j+1}-P_{H_{X}^{1,m-j}}X_{m-j+1}\right) & \mbox{otherwise}
\end{cases}.
\]
The sequence $\left(X_{m-j+1}-P_{H_{X}^{1,m-j}}X_{m-j+1}\right)_{j\in\left\{ 1\ldots m\right\} }$
is a set of orthogonal vectors obtained by a Gram-Schmidt orthonormalization
procedure. Indeed, it corresponds to the series of innovations of
the time series. This implies, by orthogonality and decomposition
on the Gram-Schmidt basis, 
\[
E\left[\left(P_{H_{X}^{1,m}}X_{m+1}\right)\left(X_{m-j+1}-P_{H_{X}^{1,m-j}}X_{m-j+1}\right)^{T}\right]=\Theta_{m,j}\Sigma_{m-j}
\]
where $\Sigma_{m-j}$ is the variance matrix of the corresponding
innovation process (perturbations to the linear model). $X_{m+1}-P_{H_{X}^{1,m}}X_{m+1}$
is orthogonal to $X_{m-j+1}-P_{H_{X}^{1,m-j}}X_{m-j+1}$ therefore
\[
E\left(X_{m+1}\left(X_{m-j+1}-P_{H_{X}^{1,m-j}}X_{m-j+1}\right)^{T}\right)=\Theta_{m,j}\Sigma_{m-j}.
\]
Substituting $P_{H_{X}^{1,m-j}}X_{m-j+1}$ by $\sum_{i=1}^{m-j}\Theta_{m-j,j}\left(X_{m-j-i+1}-P_{H_{X}^{1,m-j-i}}X_{m-j-i+1}\right)$
we get 
\[
E\left(X_{m+1}\left(X_{m-j+1}^{T}-\sum_{i=1}^{m-j}\left(X_{m-j-i+1}-P_{H_{X}^{1,m-j-i}}X_{m-j-i+1}\right)^{T}\Theta_{m-j,j}^{T}\right)\right)=\Theta_{m,j}\Sigma_{m-j}.
\]
And, therefore, 
\[
\gamma_{-j}^{X}-\sum_{i=1}^{m-j}\Theta_{m,j+i}\Sigma_{m-j-i}\Theta_{m-j,j}^{T}=\Theta_{m,j}\Sigma_{m-j}.
\]
Substituting $j$ by $m-j$ gives 
\[
\gamma_{j-m}^{X}-\sum_{i=1}^{j}\Theta_{m,m-j+i}\Sigma_{j-i}\Theta_{j,m-j}^{T}=\Theta_{m,m-j}\Sigma_{j}
\]
and finally substituting $j-i$ by $i$ yields 
\[
\gamma_{j-m}^{X}-\sum_{i=0}^{j-1}\Theta_{m,m-i}\Sigma_{i}\Theta_{j,m-j}^{T}=\Theta_{m,m-j}\Sigma_{j}.
\]
Finally, Pythagora's theorem for orthogonal projections implies that
\[
\Sigma_{m}=\mbox{Var}\left(X_{t+1}\right)-\mbox{Var}\left(P_{H_{X}^{1,t}}X_{t+1}\right)=\gamma_{0}^{X}-\sum_{i=0}^{m-1}\Theta_{m,m-i}\Sigma_{i}\Theta_{m,m-i}^{T}.
\]
Based on the estimates of $\gamma^{X}$ given by the averaging procedure
above, a recursive procedure, starting by $\widehat{\Sigma_{0}}=\widehat{\gamma_{0}^{X}}$
yields consistent estimates for $\widehat{\Sigma_{m}}$ and $\widehat{\Theta_{1}},\ldots,\widehat{\Theta}_{q}$.
In the case of an order $q$ MA model, $B_{1},\ldots,B_{q}=\Theta_{1},\ldots,\Theta_{q}$
and therefore the we get the estimates of the parameters of the model.
The complexity of the algorithm here is $O\left(p^{2}d^{3}\right)$.

\paragraph{Recursive procedure:}

One starts with $\Sigma_{0}=\widehat{\gamma}_{0}^{X}$ and then for
$m\in\left\{ 1\ldots q\right\} $ computes

\[
\forall j\in\left\{ 0\ldots m-1\right\} ,\:\widehat{\Theta}_{m,m-j}=\left[\widehat{\gamma}_{j-m}^{X}-\sum_{i=0}^{j-1}\widehat{\Theta}_{m,m-i}\widehat{\Sigma}_{i}\widehat{\Theta}_{j,j-i}^{T}\right]\widehat{\Sigma}_{j}^{-1}
\]
and 
\[
\widehat{\Sigma}_{m}=\widehat{\gamma}_{0}^{X}-\sum_{i=0}^{m-1}\Theta_{m,m-i}\Sigma_{i}\Theta_{m,m-i}^{T}.
\]

\subsection{ARMA model estimation:}

We are now concerned with estimating the parameters of 
\[
X_{t}=A_{1}X_{t-1}+\ldots+A_{p}X_{t-p}+\varepsilon_{t}+B_{1}\varepsilon_{t-1}+\ldots+B_{q}\varepsilon_{t-q}.
\]
We assume in the following that the model is causal. This implies
in particular the existence of matrices $\left(\varPsi_{j}\right)_{j=0}^{+\infty}$
such that $X_{t}=\sum_{j=0}^{\infty}\varPsi_{j}\varepsilon_{t-j}$.
As $\left(\varepsilon_{t}\right)_{t\in\mathbb{Z}}$ is a white noise
process, necessarily 
\[
\begin{cases}
\varPsi_{0} & =I\\
\varPsi_{j} & =B_{j}+\sum_{i=1}^{\min\left(j,p\right)}A_{i}\varPsi_{j-i}
\end{cases}
\]
where, by convention, $B_{j}=0$ whenever $j>q$ and $A_{j}=0$ whenever
$j>p$. By construction, $\left(\varPsi_{j}\right)_{j=1}^{p+q}$ can
be estimated thanks to the innovation estimates $\left(\widehat{\varPsi}_{p+q,j}\right)_{j=1}^{p+q}$
provided by the innovation algorithm above. Then one has 
\[
\widehat{\varPsi}_{p+q,j}=\widehat{B}_{j}+\sum_{i=1}^{\min\left(j,p\right)}\widehat{A}_{i}\widehat{\varPsi}_{p+q,j-i},\:\forall j\in\left\{ 1\ldots p+q\right\} .
\]
As by convention, $\forall j>p,\: B_{j}=0$, necessarily, the estimates
$\left(\widehat{A}_{j}\right)_{j=1}^{p}$ solve the following linear
system:
\[
\left[\begin{array}{cccc}
\widehat{\varPsi}_{p+q,\: q}^{T} & \widehat{\varPsi}_{p+q,\: q-1}^{T} & \cdots & \widehat{\varPsi}_{p+q,\: q+1-p}^{T}\\
\widehat{\varPsi}_{p+q,\: q+1}^{T} & \widehat{\varPsi}_{p+q,\: q}^{T} &  & \widehat{\varPsi}_{p+q,\: q+2-p}^{T}\\
\vdots &  & \ddots & \vdots\\
\widehat{\varPsi}_{p+q,\: p+q-1}^{T} & \widehat{\varPsi}_{p+q,\: p+q-2}^{T} & \cdots & \widehat{\varPsi}_{p+q,\: q}^{T}
\end{array}\right]\left[\begin{array}{c}
\widehat{A_{1}}^{T}\\
\vdots\\
\widehat{A}_{p}^{T}
\end{array}\right]=\left[\begin{array}{c}
\widehat{\varPsi}_{p+q,\: q+1}^{T}\\
\vdots\\
\widehat{\varPsi}_{p+q,\: p+q}^{T}
\end{array}\right]
\]
Once this block Toeplitz system has been solved, the estimates $\widehat{B}_{j}$
can be determined thanks to \textbf{
\[
\widehat{B}_{j}=\widehat{\varPsi}_{p+q,\: j}-\sum_{i=1}^{\min\left(j,p\right)}\widehat{A}_{i}\widehat{\varPsi}_{p+q,\: j-i},\:\forall j\in\left\{ 1,\ldots q\right\} .
\]
}An estimate of the diagonal noise variance matrix is given by 
\[
\widehat{\Sigma}_{p+q}=\widehat{\gamma}_{0}^{X}-\sum_{i=0}^{p+q-1}\widehat{\varPsi}_{p+q,\: p+q-i}\widehat{\Sigma}_{i}\widehat{\varPsi}_{p+q,\: p+q-i}^{T}.
\]
The complexity of the algorithm is $O\left(\left(p+q\right)^{2}d^{3}\right)$
.

\section{Predictions with linear time series models}

Linear models for time series, however often very simplistic, offer
linear predictors for the next events to occur given a series of previous
observations. Linear predictors are simple to set up and offer good
guarantees on the results they provide in the form of confidence intervals.
This principle can also be extended to that of featurized predictions
with a linear model that feeds on non-linear features. Predictions
leverage parallelism with respect to process dimensions straightforwardly
and are computed in an iterative fashion with respect to time. There
is no contribution here in that regard except in that we highlight
that predictions in the general case an ARMA process only depend on
short range previously observed values and therefore prove corresponding
computations feature weak memory.

\subsection{Predictions in the AR case}

The auto-regressive family of models is the simplest, the one step
ahead predictor of $X_{t}$, that will be denoted $\overset{\rightarrow1}{X_{t}}$
is trivially 
\[
\overset{\rightarrow1}{X_{t}}=A_{1}X_{t}+\ldots+A_{p}X_{t-p+1}
\]
and predictions more than one step ahead are obtained in a recursive
fashion by re-injecting shorter range projections into the linear
system above. The variance of predictions can then be computed based
on $\Sigma_{\varepsilon}$ (see \cite{Ltkepohl:2007:NIM:1554948} for details).

\subsection{Predictions in the ARMA case}

Moving average components introduce a supplementary difficulty in
predicting values based on passed observations. The aim of this section
is to show that predictions can be evaluated based on a recursive
procedure which, at each step, only considers previously observed
values on a short range.\\
\\
In order to forecast an ARMA process one runs the innovation algorithms
on the observations so as to obtain an estimate of the innovations.
There exists a series of projection matrices $\left(\Theta_{t,t'}\right)_{t\in\left\{ 0,T\right\} ,t'\leq t}$
such that 
\[
\begin{cases}
\forall t\in\left\{ 0,\:\max\left(p,q\right)-1\right\}  & \overset{\rightarrow1}{X_{t}}=\sum_{t'=1}^{t}\Theta_{t,t'}\left(X_{t+1-t'}-\overset{\rightarrow1}{X_{t+1-t'}}\right)\\
\forall t\in\left\{ \max\left(p,q\right),\: T\right\}  & \overset{\rightarrow1}{X_{t}}=A_{1}X_{t}+\ldots+A_{p}X_{t-p+1}+B_{1}\left(X_{t}-\overset{\rightarrow1}{X_{t-1}}\right)+\ldots+B_{q}\left(X_{t-q+1}-\overset{\rightarrow1}{X_{t-q}}\right)
\end{cases}.
\]
Therefore, at each time-step $t$, provided the innovation projection
matrices have been computed (which has been done iteratively from
$0$ to $t$), only $\max\left(p,q\right)$ observations and forecasts
need to be taken into account. This algorithm can be run in a streaming
fashion. This algorithm can be run in an approximate parallel manner
in the case of stable models in which the importance of initialization
errors decays exponentially.

\section{AR, MA, ARMA estimation by Bayesian methods}

Let us first focus on the estimation of AR models by Bayesian methods
and prove they rely on weak memory computations.

\subsection{Bayesian estimation of AR models}

At each time-step one considers an equation of the form 
\[
X_{t}=A_{1}X_{t-1}+\ldots+A_{p}X_{t-p}+\varepsilon_{t}.
\]
We assume the errors have a parametric distribution $f\left(\varepsilon_{t},\vartheta\right)=\frac{1}{Z\left(\vartheta\right)}\mbox{exp}\left(H\left(\varepsilon_{t},\vartheta\right)\right)$
where $\vartheta$ is a set of parameters in $\Theta$ (which we assume
is convex), $H$ is a function from $\mathbb{R}^{d}\times\Theta$
onto $\mathbb{R}$ and $Z$ is a function from $\Theta$ onto $\mathbb{R}^{++}$.
We further assume that $f$ is log-strongly-concave with respect to
its first argument. One will typically consider a centered Gaussian
distribution for the white noise $\varepsilon_{t}$ with variance
$\Sigma_{\varepsilon}$. In other words 
\[
f\left(\varepsilon_{t},\Sigma_{\varepsilon}\right)=\left(2\pi\right)^{-\frac{d}{2}}\det\left(\Sigma_{\varepsilon}\right)^{-\frac{1}{2}}\exp\left(-\frac{1}{2}\varepsilon_{t}^{T}\Sigma_{\varepsilon}^{-1}\varepsilon_{t}\right).
\]

\subsubsection{Iterative maximum conditional likelihood based estimation}

In this section, one assumes we are given a series of samples $\left(X_{1},\ldots,X_{N}\right)$
where $N>p$. We want to optimize the likelihood of samples considering
that the first $p$ observations are not perturbed $\left(\varepsilon_{1}=0,\:\varepsilon_{2}=0,\ldots.\:\varepsilon_{p}=0\right)$.
The likelihood function decomposes as follows: 
\[
\mathcal{L}\left(X_{1},\ldots,X_{N},\: A_{1},A_{2},\ldots,A_{p},\:\vartheta\right)=\mathcal{L}\left(X_{1},\ldots,X_{p},\: A_{1},\ldots,A_{p},\:\vartheta\right)\prod_{t=p+1}^{N}f\left(X_{t}-A_{1}X_{t-1}-\ldots-A_{p}X_{t-p},\vartheta\right)
\]
The log-likelihood is therefore 
\[
\log\mathcal{L}\left(X_{1},\ldots,X_{N},\: A_{1},A_{2},\ldots,A_{p},\:\vartheta\right)=\log\mathcal{L}\left(X_{1},\ldots,X_{p},\: A_{1},\ldots,A_{p},\:\vartheta\right)
\]
\[
+\sum_{t=p+1}^{N}\log f\left(X_{t}-A_{1}X_{t-1}-\ldots-A_{p}X_{t-p},\:\vartheta\right)
\]
Our aim here is to find $\vartheta$ and $A_{1},\ldots,A_{p}$ so
as to maximize $\log\mathcal{L}\left(X_{1},\ldots,X_{N},\: A_{1},A_{2},\ldots,A_{p},\:\vartheta\right)$.
In the conditional likelihood framework we assume $X_{1},\ldots,X_{p}$
are known without perturbations and therefore the problem comes down
to 
\[
\underset{A_{1},A_{2},\ldots,A_{p},\:\vartheta}{\max}\log\mathcal{L}\left(X_{1},\ldots,X_{p},\: A_{1},\ldots,A_{p},\:\vartheta\right)+\sum_{t=p+1}^{N}\log f\left(X_{t}-A_{1}X_{t-1}-\ldots-A_{p}X_{t-p},\:\vartheta\right)
\]
Without prior knowledge on $\left(X_{1},\ldots,X_{p}\right)$, $\left(A_{1},\ldots.A_{p}\right)$
or $\Theta$, the maximum likelihood problem can be rewritten as
\[
\underset{A_{1},A_{2},\ldots,A_{p},\:\vartheta}{\max}\sum_{t=p+1}^{N}\log f\left(X_{t}-A_{1}X_{t-1}-\ldots-A_{p}X_{t-p},\vartheta\right).
\]
The objective is strongly concave with respect $\vartheta$ and $\left(A_{1},\ldots,A_{p}\right)$
separately. An alternate maximization procedure therefore yields an
argument-wise maximum (but not necessarily a global optimum).

\subsubsection{Iterative un-conditional likelihood based estimation}

In this setting we consider $\left(X_{1},\ldots,X_{p}\right)$ as
unknown. We also restrict to the case in which $f$ is a Gaussian
distribution with variance $\Sigma_{\varepsilon}$. In such a context,
one can rewrite each $X_{t}$ as an infinite linear combination of
$\left(\varepsilon_{s}\right)_{s\leq t}$. The variables $\left(\varepsilon_{s}\right)_{s\leq t}$
are independent Gaussian vectors and therefore the infinite linear
combination is also a Gaussian variable. Let $\Gamma_{p}^{0}$ the
variance of $\left(\begin{array}{c}
X_{1}\\
\vdots\\
X_{p}
\end{array}\right)$. The problem becomes that of finding

\[
\underset{X_{1},X_{2},\ldots,X_{p},A_{1},A_{2},\ldots,A_{p},\:\vartheta}{\max}\begin{array}{c}
\frac{1}{2}\log\left(\det\left(\left(\Gamma_{p}^{0}\right)^{-1}\right)\right)-\left(X_{1}^{T},\ldots,X_{p}^{T}\right)\left(\Gamma_{p}^{0}\right)^{-1}\left(\begin{array}{c}
X_{1}\\
\vdots\\
X_{p}
\end{array}\right)\\
+\frac{N-p}{2}\mbox{log}\left(\mbox{det}\left(\left(\Sigma_{\varepsilon}^{-1}\right)\right)\right)\\-
\sum_{t=p+1}^{N}\left(X_{t}-A_{1}X_{t-1}-\ldots-A_{p}X_{t-p}\right)^{T}\left(\Sigma_{\varepsilon}\right)^{-1}\left(X_{t}-A_{1}X_{t-1}-\ldots-A_{p}X_{t-p}\right)
\end{array}.
\]
It is somewhat similar yet somewhat more complex than the conditional
likelihood based estimator.

\subsubsection{First order methods based resolution}

For strongly concave and Lipschitz gradient optimization problems,
gradient ascent yields an exponential rate of convergence. The gradient
of the conditional likelihood problem decomposes as 

\[
\sum_{t=p+1}^{N}\nabla\log f\left(\vartheta,X_{t}-A_{1}X_{t-1}-\ldots-A_{p}X_{t-p}\right).
\]
What is remarkable there is that only local data (namely $\left(X_{t},\ldots X_{t-p}\right)$
is needed to compute the term corresponding to datum $X_{t}$.\\
\\
For strongly concave and Lipschitz gradient optimization problems
consisting of a large sum, stochastic gradient descent has a squared
$L_{2}$ error that converges in $\frac{1}{\mbox{n iterations}}$
to $0$ with an appropriate hyperbolically decreasing step size. In
a big data context, if $N$ is so high that a holistically computation
of the gradient is too computationally expensive then one should use
a stochastic gradient method in order to estimate their model.\\
\\
One should definitively avoid second order methods here if $d\sim10^{4}$
and choose a deterministic or stochastic first order optimization
method. Practically this requires to sample out a certain value of
$t$ in $\left\{ p+1\ldots N\right\} $ and compute 
\[
\nabla_{A_{1},\ldots,A_{p}}\log f\left(\vartheta,X_{t}-A_{1}X_{t-1}-\ldots-A_{p}X_{t-p}\right)
\]
 and $\nabla_{\vartheta}\log f\left(\vartheta,X_{t}-A_{1}X_{t-1}-\ldots-A_{p}X_{t-p}\right)$.
There again, only local data is needed to compute the gradient contribution
corresponding to $X_{t}$.

\section{When models become very highly dimensional}

\begin{figure}
\begin{centering}
\includegraphics[width=12cm]{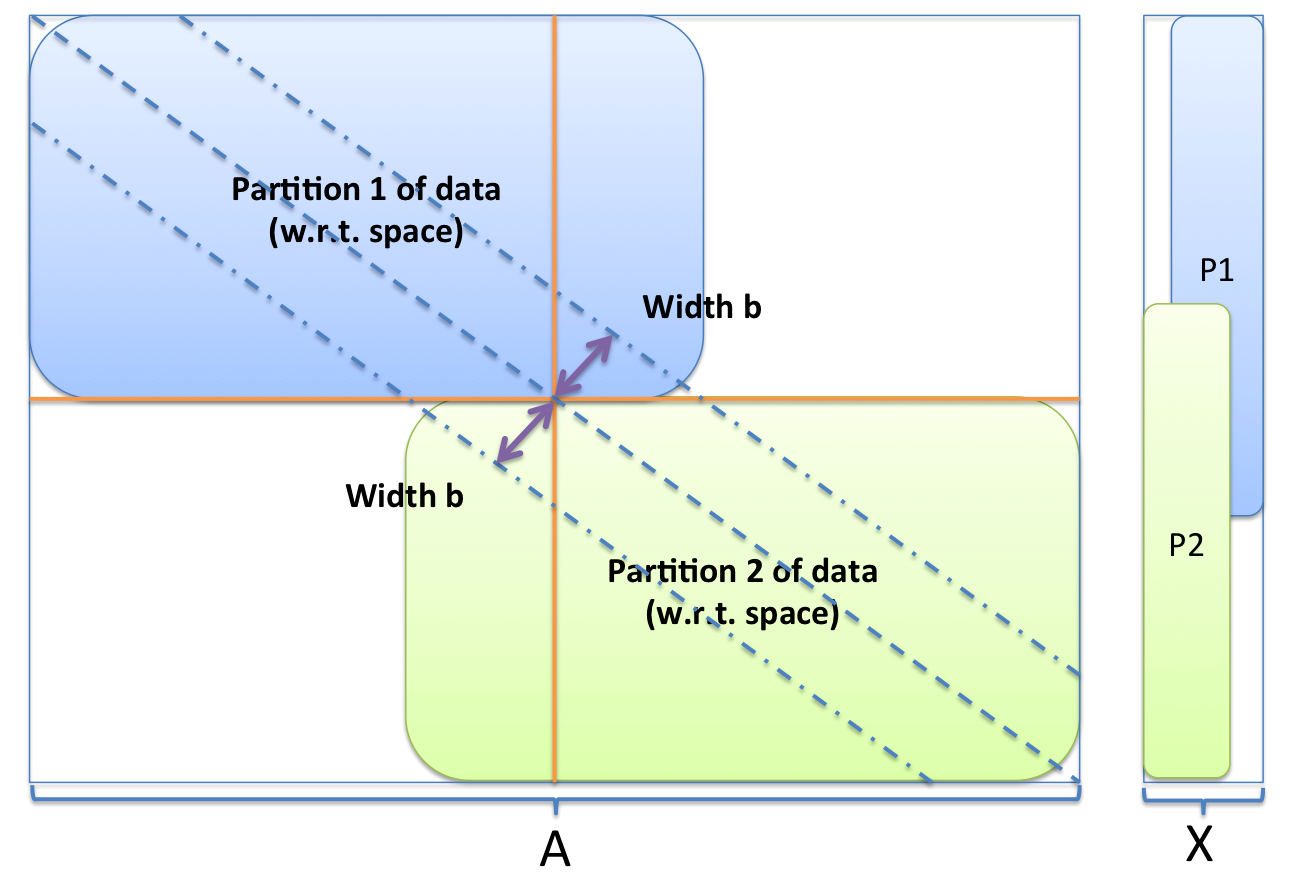}
\par\end{centering}

\protect\caption{Banded $A$ matrix.\label{fig:Banded--matrix}}

\end{figure}

Let us come back to an order $p$ auto-regressive model with $d$
spatial dimensions where $d\gg p$. The Yule-Walker equations only
yield estimates that will require the inversion of a size $d$ square
matrix, this is not a scalable solution with respect to $d$.\\
\\
Let us assume the order of the model is $1$ (one can always rewrite
an order $p$ model in this manner): 
\[
X_{t+1}=AX_{t}+\varepsilon_{t}
\]
where $A$ is sparse and rearranged so as to be a width $b$ banded
matrix with $b\ll d$. Such a sparsity pattern is quite frequent and
arises for instance in numerical differentiation schemes. It is illustrated
in Figure \ref{fig:Banded--matrix}.

\subsection{Efficient one-step-ahead prediction with spatially sparse models}

If one observes $X_{t}$ at a given timestamp $t$, then the best
linear predictor of $X_{t+1}$ is 
\[
\widehat{X}_{t+1}=AX_{t}.
\]
Let us assume we split $A$ into a row partitioned matrix as follows:
\[
A=\left[\begin{array}{ccccc}
A_{1,0} & A_{1,1}\\
A_{2,-1} & A_{2,0} & A_{2,1}\\
 & A_{3,-1} & A_{3,0} & A_{3,1}\\
 &  & \ddots & \ddots & A_{P-1,1}\\
 &  &  & A_{P,-1} & A_{P,0}
\end{array}\right].
\]
For any $i\in\left\{ 1\ldots P\right\} $ let us denote $P_{i}$ the
set of indices spanned by the rows of $A_{i,0}$ and $P_{i}^{+}$
the set of indices spanned by $A_{i,-1},\: A_{i,0},\: A_{i,1}$.\\
\\
The class of sparse models considered above is a linear specific case
belonging to the family 
\[
X_{t+1}=\left[\begin{array}{c}
A_{\vartheta}^{1}\left(X_{t}^{P_{1}^{+}}\right)\\
A_{\vartheta}^{2}\left(X_{t}^{P_{2}^{+}}\right)\\
\vdots\\
A_{\vartheta}^{P-1}\left(X_{t}^{P_{P-1}^{+}}\right)\\
A_{\vartheta}^{P}\left(X_{t}^{P_{P}^{+}}\right)
\end{array}\right]+\varepsilon_{t}
\]
where $\left(A_{\vartheta}^{i}\right)_{i\in\left\{ 1\ldots P\right\} }$
is a family of functions from $\mathbb{R}^{\left|P_{i}^{+}\right|}$
onto $\mathbb{R}^{\left|P_{i}\right|}$ parametric by $\vartheta$.\\
\\
An efficient unbiased estimator of the one-step-ahead predicted value
can be written in a row partitioned manner: 
\[
\widehat{X}_{t+1}=\left[\begin{array}{c}
\widehat{X_{t}^{P_{1}}}\\
\widehat{X_{t}^{P_{2}}}\\
\vdots\\
\widehat{X_{t}^{P_{P-1}}}\\
\widehat{X_{t}^{P_{P}}}
\end{array}\right]=\left[\begin{array}{c}
A_{\vartheta}^{1}\left(X_{t}^{P_{1}^{+}}\right)\\
A_{\vartheta}^{2}\left(X_{t}^{P_{2}^{+}}\right)\\
\vdots\\
A_{\vartheta}^{P-1}\left(X_{t}^{P_{P-1}^{+}}\right)\\
A_{\vartheta}^{P}\left(X_{t}^{P_{P}^{+}}\right)
\end{array}\right].
\]
In the linear case above, the time complexity of the operation is
$O\left(d\times\left(2b+1\right)\right)\ll O\left(d^{2}\right)$.

\subsection{Efficient sparsity leveraging Bayesian estimation}

We assume that $\left(\varepsilon_{t}\right)$ is a Gaussian white
noise whose precision matrix, $\Pi_{\varepsilon}=\Sigma_{\varepsilon}^{-1}$
is block diagonal with blocks corresponding to the sets of rows and
columns $P_{1},\: P_{2},\ldots,P_{P}$.

\[
\Pi_{\varepsilon}=\Sigma_{\varepsilon}^{-1}=\left[\begin{array}{ccccc}
\pi_{1} & 0\\
0 & \pi_{2} & 0\\
 &  & \ddots\\
 &  & 0 & \pi_{P-1} & 0\\
 &  &  & 0 & \pi_{P}
\end{array}\right].
\]
The conditional log-likelihood of an observed process is therefore
\[
\mathcal{L}\left(\vartheta,\Pi_{\varepsilon}\right)=\frac{N-1}{2}\mbox{log}\left(\mbox{det}\left(\Pi_{\varepsilon}\right)\right)-\sum_{t=1}^{N-1}\left(X_{t+1}-A_{\vartheta}\left(X_{t}\right)\right)^{T}\Pi_{\varepsilon}\left(X_{t+1}-A_{\vartheta}\left(X_{t}\right)\right).
\]
Consider 
\[
\mathcal{L}_{t}\left(\vartheta,\Pi_{\varepsilon}\right)=\left(X_{t+1}-A_{\vartheta}\left(X_{t}\right)\right)^{T}\Pi_{\varepsilon}\left(X_{t+1}-A_{\vartheta}\left(X_{t}\right)\right)=\sum_{i,j=1}^{P}\left(X_{t+1}^{P_{i}}-A_{\vartheta}^{i}\left(X_{t}^{P_{i}^{+}}\right)\right)^{T}\Pi_{\varepsilon}^{P_{i},P_{j}}\left(X_{t+1}^{P_{j}}-A_{\vartheta}^{j}\left(X_{t}^{P_{j}^{+}}\right)\right).
\]
We have assumed $\forall i\neq j,\:\Pi_{\varepsilon}^{P_{i},P_{j}}=0$,
therefore 
\[
\mathcal{L}_{t}\left(\vartheta,\Pi_{\varepsilon}\right)=\sum_{i=1}^{P}\left(X_{t+1}^{P_{i}}-A_{\vartheta}^{i}\left(X_{t}^{P_{i}^{+}}\right)\right)^{T}\pi_{i}\left(X_{t+1}^{P_{i}}-A_{\vartheta}^{i}\left(X_{t}^{P_{i}^{+}}\right)\right)
\]
and furthermore 
\[
\mathcal{L}\left(\vartheta,\Pi_{\varepsilon}\right)=\frac{N-1}{2}\mbox{log}\left(\mbox{det}\left(\Pi_{\varepsilon}\right)\right)-\sum_{i=1}^{P}\sum_{t=1}^{N-1}\left(X_{t+1}^{P_{i}}-A_{\vartheta}^{i}\left(X_{t}^{P_{i}^{+}}\right)\right)^{T}\pi_{i}\left(X_{t+1}^{P_{i}}-A_{\vartheta}^{i}\left(X_{t}^{P_{i}^{+}}\right)\right).
\]
Let us consider we want to maximize $\mathcal{L}\left(\vartheta,\Pi_{\varepsilon}\right)$
with respect to $\vartheta$, we have 
\[
\nabla_{\vartheta}\mathcal{L}\left(\vartheta,\Pi_{\varepsilon}\right)=-2\sum_{i=1}^{P}\sum_{t=1}^{N-1}\left(D_{\vartheta}A_{\vartheta}^{i}\left(X_{t}^{P_{i}^{+}}\right)\right)^{T}\pi_{i}\left(X_{t+1}^{i}-A_{\vartheta}\left(X_{t}^{P_{i}^{+}}\right)\right)
\]
where, letting $\left|\vartheta\right|$ denote the number of parameters
in the model, 
\[
D_{\vartheta}A_{\vartheta}^{i}\left(X_{t}^{P_{i}^{+}}\right)=\left[\begin{array}{ccc}
\partial_{\vartheta_{1}}A_{\vartheta}^{i}\left(X_{t}^{P_{i}^{+}}\right) & \ldots & \partial_{\Theta_{\left|\vartheta\right|}}A_{\vartheta}^{i}\left(X_{t}^{P_{i}^{+}}\right)\end{array}\right].
\]
What is remarkable with these expressions is that they enable embarrassingly
parallel computations provided one uses $P$ different nodes for any
$i\in\left\{ 1\ldots P\right\} $, node $i$ holds the data corresponding
to $\left(X_{t}^{P_{i}^{+}}\right)_{t\in\left\{ 1\ldots N\right\} }$.
This is also true if one plans on using a second order method.\\
\\
With a first order method, the time complexity of computing the gradient
for each node is $O\left(N\times\left|\vartheta\right|\times\left|P_{i}^{+}\right|^{2}\right)$
where $\left|P_{i}^{+}\right|$ the cardinal of the overlapping partition
which can be as low as $2\times b+1$. This implies that this solution
is scalable with respect to the size of the model as it leverages
the prior knowledge of the model's sparsity. To the best of our knowledge,
there is no equivalent computational result with the matrix inversion
and Yule-Walker equation based methods above.

\subsection{Gradient descent, step size and rate of convergence}

This section focuses on the maximization of the conditional likelihood
of a Gaussian AR process. In order to simplify notations we only consider
the AR1 case. The conclusions below are trivially extended to the
general case. Letting $\pi_{\varepsilon}$ the precision matrix of
the process' noise, maximizing the conditional likelihood of the process
is equivalent to maximizing 
\[
\mathcal{L}\left(A_{1},\pi_{\varepsilon}\right)=\frac{1}{N}\sum_{t=1}^{N}\sum_{t=1}^{d}\sum_{t=1}^{d}\left(X_{t}^{i}-\sum_{k=1}^{d}A_{1}^{ik}X_{t-1}^{k}\right)\pi_{\varepsilon}^{i,j}\left(X_{t}^{j}-\sum_{l=1}^{d}A_{1}^{jl}X_{t-1}^{l}\right)
\]
and 
\[
\frac{\partial\mathcal{L}}{\partial A_{1}^{i_{0},j_{0}}}\left(A_{1},\pi_{\varepsilon}\right)=-\frac{2}{N}\sum_{t=1}^{N}\sum_{j=1}^{d}X_{t-1}^{j_{0}}\pi_{\varepsilon}^{i_{0},j}\left(X_{t}^{j}-\sum_{l=1}^{d}A_{1}^{jl}X_{t-1}^{l}\right).
\]
We consider $\pi_{\varepsilon}$ has already been estimated by $\widehat{\pi_{\varepsilon}}$.
Maximizing $\mathcal{L}$ with respect to $A_{1}$ then comes down
to a gradient ascent update by $\left(\frac{\partial\mathcal{L}}{\partial A_{1}^{i_{0},j_{0}}}\left(A_{1},\widehat{\pi_{\varepsilon}}\right)\right)_{i_{0},j_{0}}$.
In several particular cases this update matrix takes a remarkable
form: 
\begin{itemize}
\item If $\widehat{\pi_{\varepsilon}}=I$, $\frac{\partial\mathcal{L}}{\partial A_{1}^{i_{0},j_{0}}}\left(A_{1},\widehat{\pi_{\varepsilon}}\right)=-\frac{2}{N}\sum_{t=1}^{N}X_{t-1}^{j_{0}}\left(X_{t}^{i_{0}}-\sum_{l=1}^{d}A_{1}^{i_{0}l}X_{t-1}^{l}\right)=-2\widehat{\mbox{Cov}}\left(\overrightarrow{X_{t}},X_{t}\right)$.
In this case the Hessian of $A_{1}\rightarrow\mathcal{L}\left(A_{1},\widehat{\pi_{\varepsilon}}\right)$
is a block diagonal matrix whose blocks are $\widehat{\mbox{Cov}}\left(X_{t},X_{t}\right)$.
Therefore finding the smallest $m$ and largest $L$ eigenvalues of
$\widehat{\mbox{Cov}}\left(X_{t},X_{t}\right)$ is sufficient to compute
$\frac{2}{m+L}$. This step size is of importance as it provably achieves
an exponential rate of convergence to the optimum in a gradient ascent.
\item If $\widehat{\pi_{\varepsilon}}$ is diagonal, $\frac{\partial\mathcal{L}}{\partial A_{1}^{i_{0},j_{0}}}\left(A_{1},\widehat{\pi_{\varepsilon}}\right)=-\frac{2}{N}\sum_{t=1}^{N}X_{t-1}^{j_{0}}\widehat{\pi_{\varepsilon}^{i_{0},j_{0}}}\left(X_{t}^{i_{0}}-\sum_{l=1}^{d}A_{1}^{i_{0}l}X_{t-1}^{l}\right)=-2\widehat{\pi_{\varepsilon}^{i_{0},j_{0}}}\widehat{\mbox{Cov}}\left(\overrightarrow{X_{t}},X_{t}\right)$.
The Hessian is then the Kronecker product of $\widehat{\pi_{\varepsilon}}$
and $\widehat{\mbox{Cov}}\left(X_{t},X_{t}\right)$ which means it
is sufficient to compute the smallest and largest eigenvalues of $\widehat{\pi_{\varepsilon}}$
and $\widehat{\mbox{Cov}}\left(X_{t},X_{t}\right)$ in order to find
a converging step size to the gradient ascent. The rate of convergence
in that case will be linear.
\item In any case, a line search method also enables convergence to the
maximum.
\end{itemize}
For the detail of these convergence rates we refer the reader to

\section{Common access patterns, fragmentation of data, distribution of computations}

The M (frequentist, average based) and Z (bayesian, maximum likelihood
based) estimators above rely on similar computational needs. Namely,
one computes the output of a kernel function evaluated on neighboring
data and then averages the results.

\subsection{Map reduce for M kernel based estimators}

\begin{figure}
\begin{centering}
\includegraphics[width=12cm]{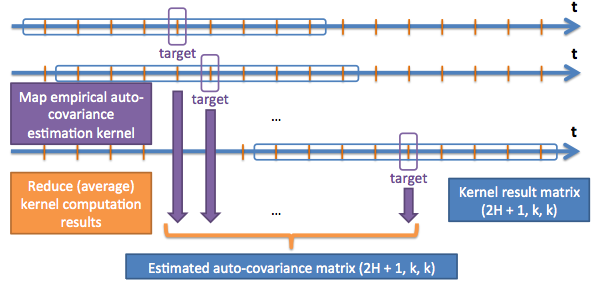}
\par\end{centering}

\protect\caption{Example of a mapped kernel based estimation in the case of the estimation
of order $H$ auto-covariance\label{fig:Autocovoriance_example}}
\end{figure}

There is a very common computation pattern to the estimators. Frequentist
estimators rely on the estimation of a covariance function. In the
centered case, this can be computed by considering a finite covariance
matrix which can be estimated by computing quantities 
\[
\widehat{\gamma_{h}^{X}}\left(\left(X_{t}\right)_{t\in\left\{ 1\ldots N\right\} }\right)=\frac{1}{N-h-1}\sum_{k=1}^{N-h}X_{k}X_{k+h}^{T}
\]
for $h\in\left\{ -H\ldots H\right\} $. This means that the vector
of square matrices 
\[
\left(\widehat{\gamma_{h}^{X}}\left(\left(X_{t}\right)_{t\in\left\{ 1\ldots N\right\} }\right)\right)_{h\in\left\{ -H\ldots H\right\} }
\]
can be approximately estimated as 
\[
\frac{1}{N-\left(2h+1\right)}\ \sum_{k=h+1}^{N-h}\left(X_{k}X_{k-h}^{T},\: X_{k}X_{k-\left(h-1\right)}^{T},\ldots,\: X_{k}X_{k}^{T},\: X_{k}X_{k+1}^{T},\ldots,X_{k}X_{k+h}^{T}\right).
\]
This estimator is biased but asymptotically unbiased. Computing this
estimator of the auto-covariance matrix is quite straightforward with
map-reduce operators. One maps the computation of the local kernel 

\[
\left(X_{k-h},\ldots,X_{k-1},X_{k},X_{k+1},\ldots,X_{k+h}\right)\overset{\kappa}{\rightarrow}\left(X_{k}X_{k-h}^{T},\: X_{k}X_{k-\left(h-1\right)}^{T},\ldots,\: X_{k}X_{k}^{T},\: X_{k}X_{k+1}^{T},\ldots,X_{k}X_{k+h}^{T}\right)
\]
and then reduces with a sum operator prior to normalizing by $\frac{1}{N-\left(2h+1\right)}$.
Such a computational schema is represented in Figure \ref{fig:Autocovoriance_example}.
This estimator is asymptotically normal with a $\frac{1}{N}$ convergence
rate.

\subsection{Map reduce for Z kernel based estimators}

Computing the gradient of parameters in an auto-regressive model of
finite order $p$ can also be formulated in terms of a similar computational
schema. For instance, when one tries to solve the conditional maximum
likelihood problem for an AR model 
\[
\underset{A_{1},A_{2},\ldots,A_{p},\:\Theta}{\max}\sum_{t=p+1}^{N}\log f\left(\Theta,X_{t}-A_{1}X_{t-1}-\ldots-A_{p}X_{t-p}\right)
\]
by a first order method, the gradient can also be computed as a sum
over a large number of terms of locally computable quantities. Indeed,
let $F$ be the function 
\[
\left(\Theta,A_{1},\ldots,A_{p}\right)\overset{F}{\rightarrow}\sum_{t=p+1}^{N}\log f\left(\Theta,X_{t}-A_{1}X_{t-1}-\ldots-A_{p}X_{t-p}\right)=\sum_{t=p+1}^{N}F_{t}\left(\Theta,X_{t}-A_{1}X_{t-1}-\ldots-A_{p}X_{t-p}\right).
\]
Then, by linearity of the gradient, 
\[
\nabla F\left(\Theta,A_{1},\ldots,A_{p}\right)=\sum_{t=p=1}^{N}\nabla F_{t}\left(\Theta,X_{t}-A_{1}X_{t-1}-\ldots-A_{p}X_{t-p}\right).
\]
For any $t\in\left\{ p+1\ldots N\right\} $, computing 
\[
\nabla F_{t}\left(\Theta,X_{t}-A_{1}X_{t-1}-\ldots-A_{p}X_{t-p}\right)
\]
only relies on data $X_{t},\ldots,X_{t-p}$.

\subsection{Overlapping data}

These quantities are straightforward to compute in a serial manner
on a single CPU. However, when it comes to using a distributed system
of computation such as Apache Spark, one has to take into account the requirement
for data to be partitioned. In order to speed up the computation process
and avoid the inter-node communication bottleneck one will want to
mostly use a data partitioning scheme that, once loaded in the RAM
of a cluster of machines, will enable embarrassingly parallel computations.\\
\\
For GPU units to speed further up the process, it is also interesting
to consider data partitioning as global memory units are usually slower
to access than shared one. Furthermore, one will want to be able to
run the computations above in an embarrassingly parallel way once
appropriate parallelization steps have been undertaken.\\
\\
As we explain in the following, this preparation step consists of
replicating the data so as to create overlapping partitioning. The
next part will explain the theoretical foundations of the scheme and
how to choose its padding horizon parameter appropriately with respect
to the calibration of a certain model. In particular, we introduce
the fundamental data structure that has been designed to enable large
scale analysis of time series: distributed overlapping blocks.

\part{A general programming paradigm for weak memory estimation}

In this paragraph we will give formal definitions of the notions of
weak memory from a computational standpoint. The following enables
general purpose time series analysis calculus to run on partitioned
data architecture in an embarrassingly parallel manner. In particular,
the distributed overlapping block data structure enables partitioning
with respect to both the different dimensions of a time series (different
signals, sources of data) and with respect to time, which is new to
the best of our knowledge. The paradigm is simple and therefore can
easily be adapted to other systems. This non system specific principle
of overlapping blocks is very powerful though as it enables in-RAM
computations on time series with an EC2 cluster in an embarrassingly
parallel way as well as GPU parallelization of time series calculus
thanks to CUDA. \\
\\
The package created for Apache Spark, SparkGeoTS, that follows this scheme
enables time series analysis at an unprecedented scale, both in terms
of density of time stamps along the time axis and in terms of very
highly dimensional time series analysis. It supports data analytics
for both regularly time stamped data and irregularly spaced time series.
Its novelty stems from leveraging the informational properties of
weak memory in ergodic time series. It is not adapted to the non-ergodic
context which is not an actual shortcoming as in this case most usual
estimators for time series are not even convergent.

\section{Weak memory computation for estimators}

The first part of this document reviewed a vast class of time
series models that feature weak memory computational needs. In the
following we formalize that notion. Let us consider data in the form
of $\left(X_{t}\right)_{t\in\left\{ 0\ldots N\right\} }$.

\paragraph{Definition: Order $H$ weak memory estimator}

An estimator $\mbox{Est}\left(\left(X_{t}\right)_{t\in\left\{ 0\ldots N\right\} }\right)$
features order $H$ weak memory if there exists an integer $H$ (horizon
in number of steps) such that sufficient statistics for that estimator
can be computed by reducing kernel computation results feeding on
a data point and the neighbors of that point that are less than $H$
time steps away on the time index.

\paragraph{Example: Second order stationary time series}

We have shown above that usual estimators for AR$(p)$ models are $p$
weak memory estimators, similarly with MA$(q)$ ($q$ weak memory) and
ARMA$(p, q)$ models ($p+q$ weak memory).

\section{Weak memory in time series graphs}

\begin{figure}
\begin{centering}
\includegraphics[width=12cm]{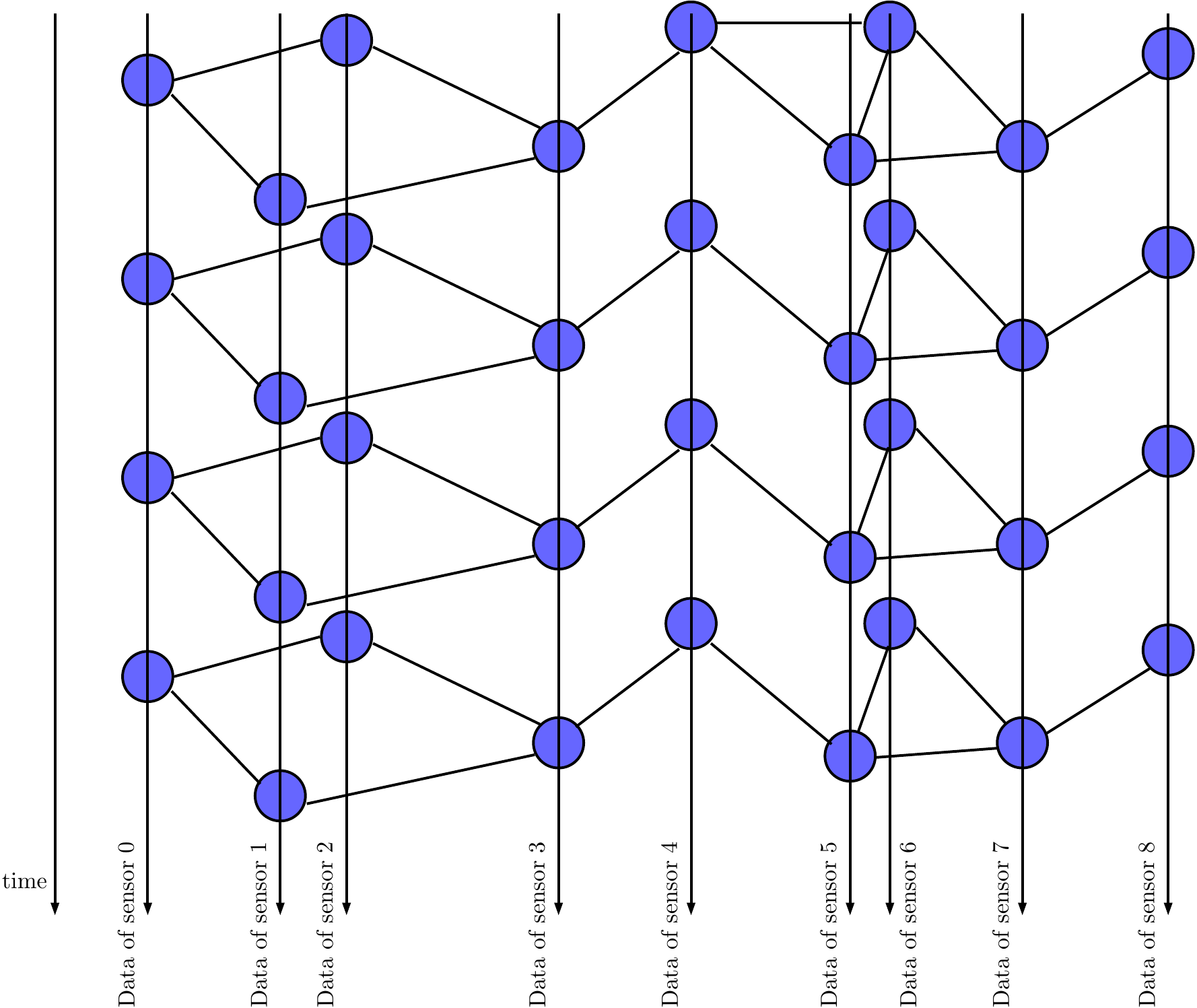}
\par\end{centering}

\protect\caption{Sensor graph lattice\label{fig:Sensor-graph-lattice}}
\end{figure}

If one considers a system in which sensors are included in a relational
graph then the time series of readings produced by these sensors are
naturally embedded in a graph. An example of the resulting data lattice
is represented in Figure \ref{fig:Sensor-graph-lattice}.\\
\\
A regularly indexed time series graph is a sequence $\left(\left(X_{t}^{v}\right)_{v\in V}\right)_{t\in\left\{ 1\ldots N\right\} }$
where the set $V$ of vertices of the graph is tied together by a set
$E$ of edges with uniform weight $1$.

\subsection{Data with regularly indexed timestamps}

\paragraph{Definition: Order $\left(H,K\right)$ weak memory estimator}

An estimator $\mbox{Est}\left[\left(\left(X_{t}^{v}\right)_{v\in V}\right)_{t\in\left\{ 1\ldots N\right\} }\right]$
is said to feature order $\left(H,K\right)$ weak memory if sufficient
statistics can be computed by reducing kernel computations that, for
each vertex state $X_{t}^{v}$, only feed on the states of neighbors
that are at most $K$ hops away in the graph on a time window that
does not go further than $H$ time steps away from $t_{0}$.

\paragraph{Example: queuing model for arterial traffic}

A traffic network can be mapped to a graph where vertices represent
intersections and edges correspond to roads. Such a mapping correspond
to a primal whose dual graph consists of vertices modeling roads and
edges accounting for the binding intersections represent. One considers
a discrete model in which the state of congestion at time $t$ on
vertex $v$ only depends on the state of congestion of its upstream
and downstream neighbors. This assumption is very usual in traffic
as soon as the discretization resolution is such that $\frac{\Delta x}{\Delta t}\leq v$
where $\Delta x$ is the length of the shortest road, $\Delta t$
the time discretization resolution and $v$ the maximum speed of vehicles
in the network.

\section{Embarrassingly parallel weak memory time series distributed representations}

In this section we define a programming paradigm to leverage the informational
structure of short memory time series in order to make their analysis
distributed and embarrassingly parallel. In particular we focus on
the case where both the sampling rate and the number of dimensions
of the time series under consideration are high enough so that the
data cannot fit in a reasonable amount of RAM on a single machine.

\subsection{Very high dimensional short memory time series}

Strategies have been developed to parallelize the analysis of time
series on distributed clusters or high performance computers. For
Apache Spark, libraries such as SparkTS and Thunder divide the time series
into smaller data sets by splitting it along dimensions. This paradigm
is efficient whenever each dimension has few enough timestamp for
its entire data to fit a node's RAM. Also, for multivariate analysis,
it is more suitable to preserve data locality across dimensions. The
statistical estimation code that should be used then needs not be
aware of the partitioning scheme. What is more, there is a gain in
terms of efficiency as joins on timestamps are not necessary.\\
\\
In the following we devise data structures that leverage the informational
properties of short memory time series in order to make their analysis
embarrassingly parallel. From a design standpoint, a map-reduce programming
scheme is adopted so that pre-existing estimation tools can be used
with that distributed container.

\subsection{The overlapping data model}

Going back to the definition of short memory in time series, it is
obvious now that any local operation only needs to be aware of its
$H$ neighborhood. Neighborhoods are defined in terms of steps for
regularly spaced data and units of time for irregularly spaced data.\\
\\
Considering the data of a time series $\left(t_{i},X_{t_{i}}\right)_{i\in\left\{ 1\ldots kN\right\} }$
and an estimation kernel $k\left(t_{0},\left(t_{j},X_{t_{j}}\right)_{\left|t_{j}-t_{0}\right|\leq H}\right)$
(with window width $2\times H$), any estimation based on a reduced
quantity of the results of this kernel can become embarrassingly parallel
provided data is partitioned, partially replicated and contained in
an overlapping partitioning with overlap at least $H$.

\subsubsection{Map-reduce based estimation with a kernel of width $2\times H$}

\begin{figure}
\begin{centering}
\includegraphics[width=12cm]{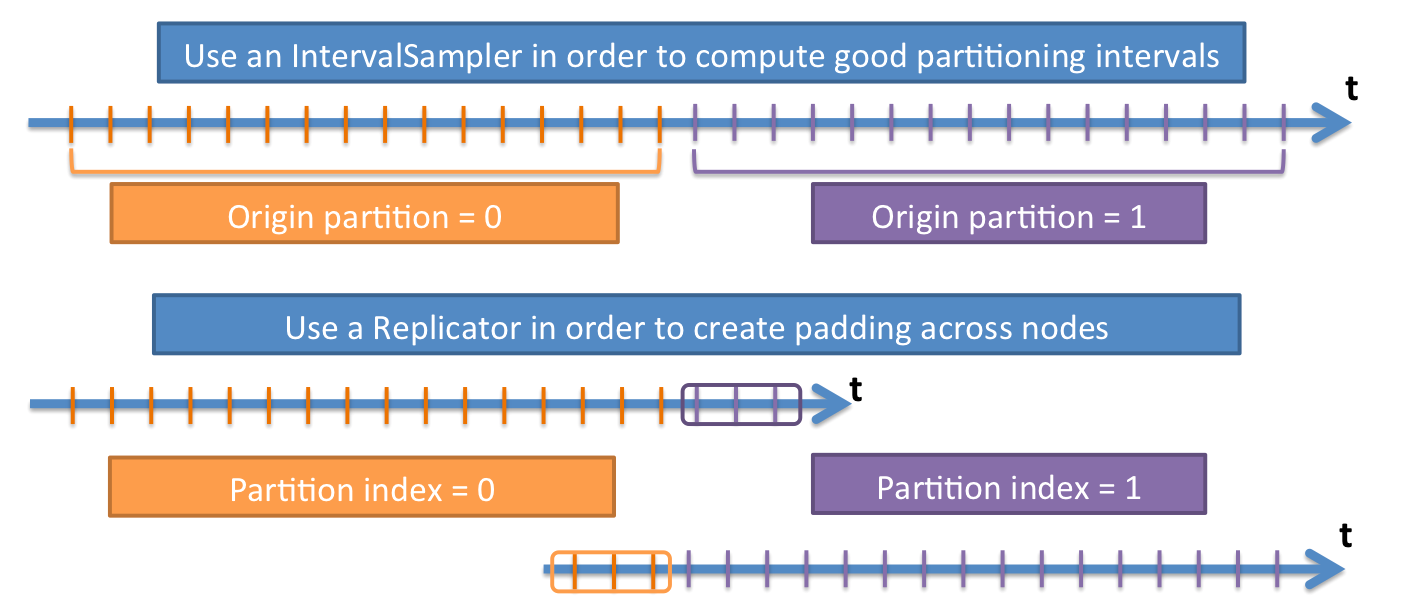}
\par\end{centering}

\protect\caption{Overlapping distributed dataset.\label{fig:Overlapping-distributed-dataset}}
\end{figure}

Let an estimator 
\[
\sum_{i=1}^{N}k\left(t_{0},\left(t_{j},X_{t_{j}}\right)_{\left|t_{j}-t_{0}\right|\leq H}\right)
\]
where $\sum$ stands for any commutative and associative operation.\\
\\
Let us assume that the data has been partitioned in $k$ partitions
with an overlap of width $H$ between partitions. The computation
flow is illustrated in the case of the estimation of an auto-covariance
function in Figure \ref{fig:Autocovoriance_example}. The presence
of an overlap directly enables one to make that computational embarrassingly
parallel with no communication needed between computational nodes
holding different partitions of the data. Such a representation of
data is illustrated in Figure \ref{fig:Overlapping-distributed-dataset}.

\subsubsection{Overlapping data set with respect to space}

\begin{figure}
\begin{centering}
\includegraphics[width=12cm]{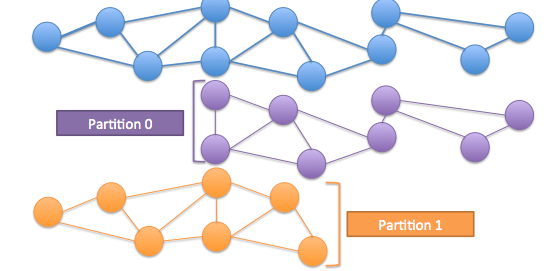}
\par\end{centering}

\protect\caption{Overlapping graph structure\label{fig:Overlapping-graph-structure}}
\end{figure}

Time is not the only dimension along which overlapping partitioning
can be computed. Obviously the scheme can be adapted to spatial meshes
with regular indexing such as images. The case of data points irregularly
arranged on a map falls under the same paradigm. More interesting
is the case of data where no Euclidian distance is available in a
straightforward way. Graphs correspond to that kind of data.\\
\\
Local operations on graphs are straightforwardly defined in terms
of reduced kernels provided they are only interested in the parenthood
degree of the neighbors of the target. Therefore the same scheme can
be adapted as illustrated in Figure \ref{fig:Overlapping-graph-structure}.

\subsection{Long memory case}

Time series such as stock valuations on the stock market or volatility
are known to feature long memory. In other words, the informational
footprint of an event never completely fades away in the system. After
having computed its consecutive differences, any integrated process
comes down to a short memory time series. This means that the data
representation paradigm above is valid in a vast range of cases. It
can be extended to partially integrated processes thanks to partial
differentiation provided the partial differentiation kernel is approximated
by a finite support kernel.

\section{Embarrassingly parallel representation of time series embedded in
graphs}

This section will be dedicated to finding applications of the graph
overlapping paradigm above in vast systems where information flows
within a sparse Dynamic Bayesian Network.

\subsection{Sparse spatial dynamic bayesian network}

Studying arterial network dynamics in traffic often comes down to
considering the series of states of vertices (road links) bound together
by intersections. In order to compute the current number of vehicles
on a link, one takes into account the number of vehicles that are effectively
leaving, baring the constraints of downstream occupancy capacity,
to the current occupancy and adds up the number of vehicles flowing
from upstream.

\subsubsection{Order $\left(1,1\right)$ time and space memory Directed Acyclic
Bayesian network}

\begin{figure}
\begin{centering}
\includegraphics[width=12cm]{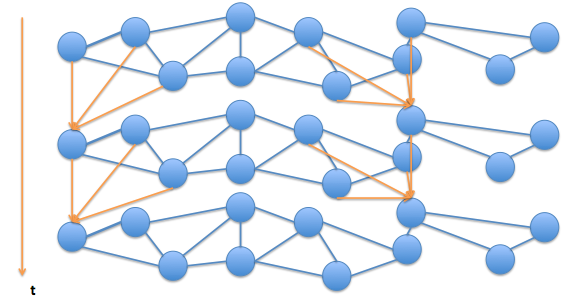}
\par\end{centering}

\protect\caption{Order $\left(1,1\right)$ time/space memory Bayesian network (for
a discretized arterial traffic model)\label{fig:Order-1-1}}
\end{figure}

Such a setting is a particular instantiation of a short time and space
memory time series graph. It is illustrated in Figure \ref{fig:Order-1-1}.\\
\\
In such a framework, the next state of a given vertex can be computed
as the result of the convolution of a kernel from its direct parents.
Therefore, the overlapping data structure can be used in order to
make the corresponding study embarrassingly parallel. This can be
used for retrospective data analysis and simulation as well provided
the same random number series are provided to edge representing identical
forward state computations.

\subsection{Cross-product of overlapping representations}

\begin{figure}
\begin{centering}
\includegraphics[width=12cm]{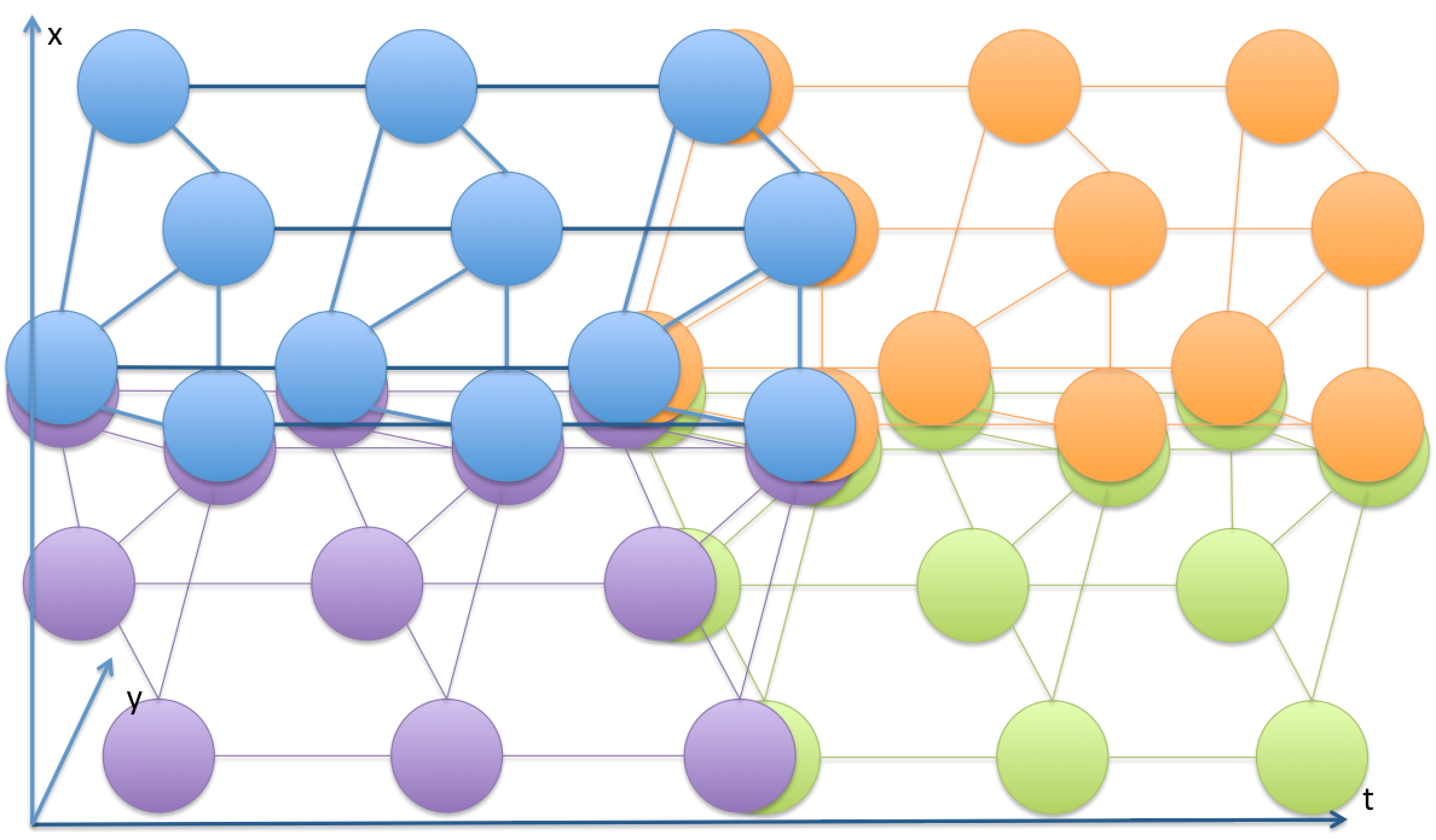}
\par\end{centering}

\protect\caption{Cross product of overlapping partitioning (with respect to time and
space)\label{fig:Cross-product}}

\end{figure}

In order to maintain reasonable size partitions, that is to say, partitions
that can fit in the RAM of a non HPC machine, one can combine overlapping
partitioning with respect to time and the corresponding time series
graph in a cross-product fashion. This creates more redundant data
but still provides a representation of data that enables embarrassingly
parallel computations if short memory is leveraged.\\
\\
If the informational structure of the data set is as presented in
Figure \ref{fig:(1,1) time series graph}, then the cross product
partitioning illustrated in Figure \ref{fig:Cross-product} enables
its embarrassingly parallel analysis.

\begin{figure}
\begin{centering}
\includegraphics[width=12cm]{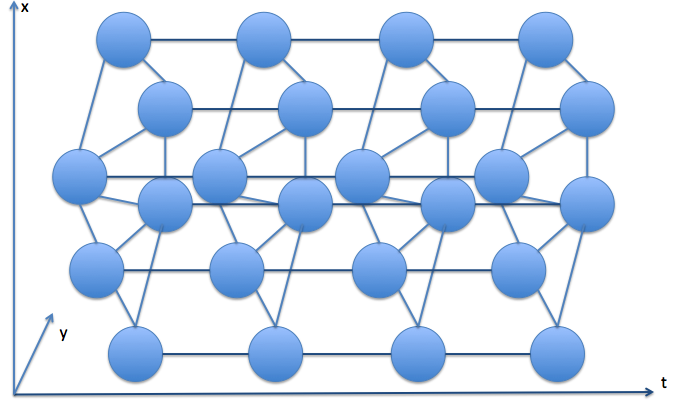}
\par\end{centering}

\protect\caption{Data embedded in a $\left(1,1\right)$ memory time series graph\label{fig:(1,1) time series graph}}

\end{figure}

\section{Speeding up computations on a GPU}

Here we prove that the overlapping block scheme is not system specific
and can be adapted to another kind of hardware: Graphical Processing
Unit (SIMD systems).

\subsection{Memory hierarchy}

Here we focus on the memory hierarchy highlighted by Nvidia's CUDA
GPGPU language. It is possible to write data from the RAM to both
the global device memory and shared memory.\\
\\
The shared memory is only accessible by threads of the same block
and only lives as long as a kernel execution. However, it enables
one to leverage the computational power of the GPU as its bandwidth
is often $100$ times that of the global memory.\\
\\
Time series analysis relies on computation of local kernels in which
data is mostly accessed redundantly by several threads of the same
block. Therefore, it seems reasonable to try and leverage this shared
memory. Each block of this memory is only accessible to the corresponding
threads and the size is much less than that of the global memory (by
a factor of $1000$).\\
\\
We show here that overlapping blocks enable us to conduct time series
analysis in a weak memory context in an embarrassingly parallel manner
while leveraging the high bandwidth of shared GPU memory.

\subsection{Parallel computation and overlapping blocks in shared memory for
regularly spaced time series}

\begin{figure}
\begin{centering}
\includegraphics[width=12cm]{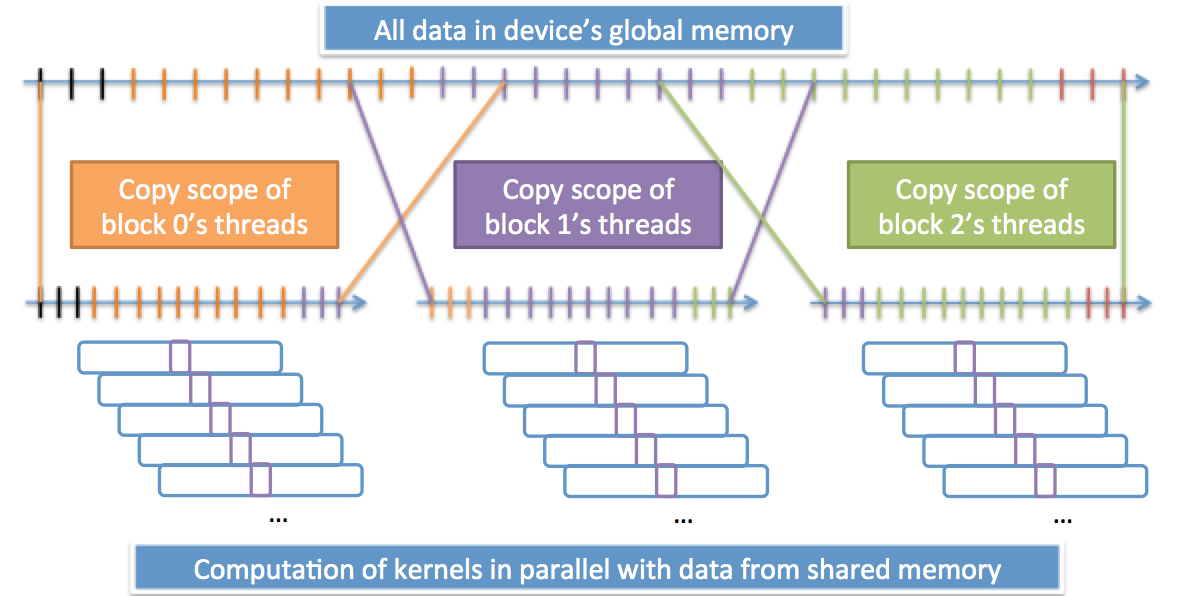}
\par\end{centering}

\protect\caption{Overlapping block enabling shared memory used on a GPU\label{fig:CUDA}}

\end{figure}

In the following we illustrate the use of overlapping blocks on GPUs
with regularly spaced time series.\\
\\
We assume the data $\left(X_{t}\right)$ has been copied in the global
GPU memory and one wants to execute a kernel of width $2H$ prior
to conducting a reduction on the results.\\
\\
Let $T_{i,j}$ the $j^{th}$ thread of the $i^{th}$ thread block.
We assume blocks are of size $N_{B}+2H$ where $N_{B}$ is small enough
so the shared memory of each block is not saturated. $T_{i,j}$ will
copy $X_{i\left(N_{B}\right)+j}$ from the global memory (address
$iN_{B}+j$) to the shared memory (local address $j$).\\
\\
Then each thread with index $j\in\left\{ H,\: N_{B}+H\right\} $ will
compute a kernel based on data contained in local addresses corresponding
to the appropriate data, namely that contained in the shared memory
addresses ranging from $j-H$ to $j+H$.\\
\\
Provided the kernel width outweighs the cost of the copy from the
global device memory to the block's shared memory, this provides a
speed up and optimized embarrassingly parallel data analysis for weak
memory systems. The flow of computations is represented in Figure
\ref{fig:CUDA}.

\subsection{Parallel computation and overlapping blocks in shared memory for
irregularly spaced time series}

For irregularly spaced data, computations are slightly more challenging
as the data that needs to be taken into account by the computation
of a local kernel does not straightforwardly correspond to a range
of memory addresses. However, we make a weak memory assumption that
no data further away than $H$ needs to taken into account in order
to compute a kernel around a datum.\\
\\
One creates an overlapping partitioning of the data with $H$ overlap
and baring the memory size constraint of the shared memory blocks.
If the kernel considers all data points with timestamps in $\left[t-H,t+H\right]$
in order to compute its output about a datum at $t$, then letting
each thread assigned with a kernel computation look backward and forward
from $t$ for the first indices whose timestamps are out of range
and then run the kernel convolution on the data within range solves
the problem.\\
\\
When kernel centers do not necessarily to data points within the data
set under consideration, things get more complicated but can be made
computationally efficient thanks to binary search. One may also create
a skip list in the global memory of the device for each thread to
access and figure out the range of the data it needs to consider.
This skip list of timestamps may sit in the global memory as it will
be accessed only once by each thread. Copying it to shared memory
blocks would cost too much overhead.

\section*{Conclusion}

In this document we have shown how weak memory models in time series
analysis can be estimated and used in the context of big distributed
data sets. Identifying how many lagged values are necessary to the
calibration of the model the user wants to implement is a necessary
preliminary step. It paves the way to building up a distributed set
of overlapping partitions. This overlapping partition scheme corresponds
to partitioning with respect to time which is the new contribution
the present document presents. We show how this can also be extended
to spatial partitioning when banded causal relationship models are
being considered.\\
The new overlapping distributed data set presented here enables a
new any-scale any-dimensional analysis of data without the need
for shuffling observations between computation nodes once the appropriate
data representation has been created. This provides the user with
the opportunity to calibrate linear weak memory time series at scale
in a reactive manner and the possibility to quickly assess how much
a given model is appropriate in terms of goodness of fit and complexity.
This paradigm can also be extended to the calibration of conditionally
heteroscedastic models \cite{lund2006estimation} as well as long
memory models \cite{doukhan2003theory,mandelbrot2013fractals}. This
is the subject of ongoing work.


\begin{thebibliography}{10}

\bibitem{brillinger1981time}
D.~R. Brillinger, {\em Time series: data analysis and theory}, vol.~36.
\newblock Siam, 1981.

\bibitem{Brockwell:1986:TST:17326}
P.~J. Brockwell and R.~A. Davis, {\em Time Series: Theory and Methods}.
\newblock New York, NY, USA: Springer-Verlag New York, Inc., 1986.

\bibitem{hamilton1994time}
J.~D. Hamilton, ``Time series analysis princeton university press,'' {\em
  Princeton, NJ}, 1994.

\bibitem{harvey1993time}
A.~C. Harvey and A.~Harvey, {\em Time series models}, vol.~2.
\newblock Harvester Wheatsheaf New York, 1993.

\bibitem{Ltkepohl:2007:NIM:1554948}
H.~Ltkepohl, {\em New Introduction to Multiple Time Series Analysis}.
\newblock Springer Publishing Company, Incorporated, 2007.

\bibitem{dean2008mapreduce}
J.~Dean and S.~Ghemawat, ``Mapreduce: simplified data processing on large
  clusters,'' {\em Communications of the ACM}, vol.~51, no.~1, pp.~107--113,
  2008.

\bibitem{shvachko2010hadoop}
K.~Shvachko, H.~Kuang, S.~Radia, and R.~Chansler, ``The hadoop distributed file
  system,'' in {\em Mass Storage Systems and Technologies (MSST), 2010 IEEE
  26th Symposium on}, pp.~1--10, IEEE, 2010.

\bibitem{thusoo2010hive}
A.~Thusoo, J.~S. Sarma, N.~Jain, Z.~Shao, P.~Chakka, N.~Zhang, S.~Antony,
  H.~Liu, and R.~Murthy, ``Hive-a petabyte scale data warehouse using hadoop,''
  in {\em Data Engineering (ICDE), 2010 IEEE 26th International Conference on},
  pp.~996--1005, IEEE, 2010.

\bibitem{zaharia2012resilient}
M.~Zaharia, M.~Chowdhury, T.~Das, A.~Dave, J.~Ma, M.~McCauley, M.~J. Franklin,
  S.~Shenker, and I.~Stoica, ``Resilient distributed datasets: A fault-tolerant
  abstraction for in-memory cluster computing,'' in {\em Proceedings of the 9th
  USENIX conference on Networked Systems Design and Implementation}, pp.~2--2,
  USENIX Association, 2012.

\bibitem{tsay2005analysis}
R.~S. Tsay, {\em Analysis of financial time series}, vol.~543.
\newblock John Wiley \& Sons, 2005.

\bibitem{mudelsee2010climate}
M.~Mudelsee, ``Climate time series analysis,'' {\em Atmospheric and}, vol.~397,
  2010.

\bibitem{basu1996time}
S.~Basu, A.~Mukherjee, and S.~Klivansky, ``Time series models for internet
  traffic,'' in {\em INFOCOM'96. Fifteenth Annual Joint Conference of the IEEE
  Computer Societies. Networking the Next Generation. Proceedings IEEE},
  vol.~2, pp.~611--620, IEEE, 1996.

\bibitem{hipel1994time}
K.~W. Hipel and A.~I. McLeod, {\em Time series modelling of water resources and
  environmental systems}.
\newblock Elsevier, 1994.

\bibitem{hunter2011scaling}
T.~Hunter, T.~Moldovan, M.~Zaharia, S.~Merzgui, J.~Ma, M.~J. Franklin,
  P.~Abbeel, and A.~M. Bayen, ``Scaling the mobile millennium system in the
  cloud,'' in {\em Proceedings of the 2nd ACM Symposium on Cloud Computing},
  p.~28, ACM, 2011.

\bibitem{franklin2013mllib}
M.~Franklin {\em et~al.}, ``Mllib: A distributed machine learning library,''
  {\em NIPS Machine Learning Open Source Software}, 2013.

\bibitem{durbin2012time}
J.~Durbin and S.~J. Koopman, {\em Time series analysis by state space methods}.
\newblock No.~38, Oxford University Press, 2012.

\bibitem{boyd2004convex}
S.~Boyd and L.~Vandenberghe, {\em Convex optimization}.
\newblock Cambridge university press, 2004.

\bibitem{bertsekas1999nonlinear}
D.~P. Bertsekas, ``Nonlinear programming,'' 1999.

\bibitem{callier2012linear}
F.~M. Callier and C.~A. Desoer, {\em Linear system theory}.
\newblock Springer Science \& Business Media, 2012.

\bibitem{kantz2004nonlinear}
H.~Kantz and T.~Schreiber, {\em Nonlinear time series analysis}, vol.~7.
\newblock Cambridge university press, 2004.

\bibitem{fan2003nonlinear}
J.~Fan and Q.~Yao, {\em Nonlinear time series: nonparametric and parametric
  methods}.
\newblock Springer Science \& Business Media, 2003.

\bibitem{Straumann05estimationin}
Straumann and T.~Mikosch, ``Estimation in conditionally heteroscedastic time
  series models,'' tech. rep., Lecture Notes in Statist. 181, 2005.

\bibitem{akaike}
H.~Akaike, ``Block toeplitz matrix inversion,'' {\em SIAM Journal on Applied
  Mathematics}, vol.~24, no.~2, pp.~234--241, 1973.

\bibitem{lund2006estimation}
R.~Lund, ``Estimation in conditionally heteroscedastic time series models,''
  {\em Journal of the American Statistical Association}, vol.~101, no.~475,
  p.~1319, 2006.

\bibitem{doukhan2003theory}
P.~Doukhan, G.~Oppenheim, and M.~S. Taqqu, {\em Theory and applications of
  long-range dependence}.
\newblock Springer Science \& Business Media, 2003.

\bibitem{mandelbrot2013fractals}
B.~B. Mandelbrot, {\em Fractals and Scaling in Finance: Discontinuity,
  Concentration, Risk. Selecta Volume E}.
\newblock Springer Science \& Business Media, 2013.

\end{thebibliography}
\end{document}